\pdfoutput = 1
\documentclass[
aps,%
12pt,%
final,%
notitlepage,%
oneside,%
onecolumn,%
nobibnotes,%
nofootinbib,%
superscriptaddress,%
noshowpacs,%
centertags]%
{revtex4}

\begin{document}

\title{Restoration of the Parameters of a Gas-Dust Disk Based on Its Synthetic Images}

\author{\firstname{A.~M.}~\surname{Skliarevskii}}
\email{sklyarevskiy@sfedu.ru}
\affiliation{%
Southern Federal University, Rostov-on-Don, Russia
}%

\author{\firstname{Ya.~N.}~\surname{Pavlyuchenkov}}
\email{pavyar@inasan.ru}
\affiliation{%
Institute of Astronomy, Russian Academy of Sciences, Moscow, Russia
}%

\author{\firstname{E.~I.}~\surname{Vorobyov}}
\affiliation{%
Institute of Astronomy, Russian Academy of Sciences, Moscow, Russia
}%
\affiliation{%
Department of Astrophysics, Vienna University, Vienna, Austria
}%


\begin{abstract}
{The topic of the present study is combining a dynamic model of a protoplanetary disk with the computations of radiation transfer for obtaining synthetic spectra and disk images suitable for immediate comparison of the model with observations. Evolution of the disk was computed using the FEOSAD hydrodynamic model, which includes a self-consistent calculation of the dynamics of dust and gas in the 2D thin disk approximation. Radiation transfer was simulated by the open code RADMC-3D. Three phases of disk evolution were considered: a young gravitationally unstable disk, a disk during an accretion luminosity burst, and an evolved disk. For these stages, the influence of various processes upon the disk's thermal structure was analyzed, as well as the differences between the temperatures obtained in the initial dynamic model and in the model with a detailed calculation of the radiation transfer. It is shown that viscous heating in the inner regions and adiabatic heating in the disk spirals can be important sources of heating. On the basis of the calculated spectral energy distributions, using SED-fitter software package used for the observations, physical parameters of the model disks were reconstructed. A significant spread between reconstructed parameters and initial characteristics of the disk indicates verification necessity of the models within the framework of spatially resolved observations of disks in the different spectral ranges}
\end{abstract}

\maketitle

\section{INTRODUCTION}
Formation of stars and planets is both one of the key problems in astrophysics and one of the most rapidly developing areas of observational astronomy. The stars are formed as a result of the gravitational collapse of compact gas-dust prestellar condensations. Before reaching the protostar, an essential fraction of the matter of pre-stellar condensation forms a circumstellar disk due to the angular momentum conservation of accreting matter. In this disk, the planets can form as a result in the following ways: the sticking of the dust particles and their subsequent gravitational association ~\cite{1996Icar..124...62P}; gravitational instability of the gaseous disk itself~\cite{2003ApJ...599..577B,2013A&A...552A.129V} or a combination of clumps formation, their migration, and degassing processes~\cite{2010MNRAS.408L..36N,2010Icar..207..509B}.

Despite the success of clarifying the formation of stars and planets, the details of the evolution of the protoplanetary disk and the role of various physical processes have not been finally clarified as yet (see review ~\cite{2011ARA&A..49...67W}). To a large extent, this is due to the difficulties of observations of protoplanetary disks (PDs). For many years, one of the main sources of information on PDs has been the analysis of the spectral distribution ofthe radiation energy of the entire disk (see,e.g.,~\cite{2007MNRAS.378..369N,2010ApJ...712..925C}). Specifically, spectral features (so-called infrared excesses) formed the basis for classification of young stellar objects~\cite{1987IAUS..115....1L}.
Relatively simple analytical models make it possible to restore disk parameters (mass, size, density, distribution profile) based on its spectrum~\cite{1990AJ.....99..924B,1997ApJ...490..368C}. However, the recovery of information about the disk structure from its integral spectrum is a complex inverse problem, with a solution that can be ambiguous~\cite{2015EPJWC.10200007W}. 

With advent of spatially resolved images of the disks at different wavelengths, it became possible to study their structure directly. While optical and infrared images of the disks, obtained by the Hubble~\cite{1994ApJ...436..194O} and
VLT~\cite{2018ApJ...863...44A}, telescopes, reveal the surface layers of the disk, radio interferometry provides information about the inner structure of the disk (see, e.g.,~\cite{1996A&A...309..493D}).
A revolutionary step in the observational study of PDs happened after the start of observations by the ALMA interferometer (see Section 4) within the framework of the DSHARP project~\cite{2018ApJ...869L..41A}. In general, PD images in different frequency ranges demonstrated the following: a variety of morphologic features of disks (rings, spirals, cavities); strong radial and vertical gradients of molecular abundances associated with the changes in physical conditions; and spectral index gradients, probably due to the evolution and migration of the dust, etc. Thus, this all provides evidence on the limitations of simple models (based on "standard" assumptions of monotonous distribution of surface density, constant dust to gas mass ratio, etc.) for interpretation of observations. At the same time, new observational data provide a basis for the development of detailed theoretical models of the evolution of disks and the study of key processes in them.

Impressive steps have already been made towards the development of detailed theoretical disk models and interpretation of observations based on them. For instance, a three-dimensional MHD model of a protoplanetary disk, together with calculation of the radiation transfer, was used to visualize disks, the evolution of which is controlled by the magneto-rotational instability (MRI) ~\cite{2015A&A...574A..68F}. In~\cite{2015ApJ...809...93D},
a three-dimensional hydrodynamic model was combined with the calculation of radiation transfer to study the observational manifestations of the gaps cleared by planets in the circumstellar disks. Combination of thermo-chemical modeling of a disk with the radiation transfer, as well as a detailed review of self-consistent modeling of the structure and images of the disks can be found in~\cite{2019PASP..131f4301W}.

However, despite the progress in this area, bringing detailed dynamic models of PDs evolution to synthetic images, their subsequent analysis and comparison with real observed images is still a fairly unique practice. Such models make it possible not only to study the manifestations of various physical processes and phenomena, but also can be used to assess the applicability of relatively simple approaches for the reconstruction of the parameters of objects using available observations. However, visualization of dynamic models in the form of synthetic observations is a difficult task, since it needs to consider many factors affecting the thermal structure of the disk and its final images. Many of these factors have not been studied adequately. In this regard, the main objectives of the present study are: 
\begin{enumerate}
\item Construction of synthetic images of PDs in various spectral ranges based on a self-consistent hydrodynamic model computed with account of a large number of key physical processes and including a self-consistent calculation of the dynamics of dust and gas. The analysis of the manifestations of inhomogeneous structures in the model disks, such as spirals and gas and dust clumps.

\item  Modeling of the thermal structure of the protoplanetary disk using a sophisticated radiation transfer method and comparison of obtained distributions with the results of hydrodynamic calculations, in which the disk temperature was calculated approximately. The analysis of discovered differences and identification of the factors affecting the temperature characteristics of the disk.

\item 
Reconstruction of the physical characteristics of PDs (mass, radius, age, etc.) from the obtained synthetic images applying the methods used in the interpretation of observational data. Comparison of the obtained characteristics of the disk with the original ones and deriving conclusions about the relevance of the approaches used to restore the parameters of the disks.
\end{enumerate}

\begin{figure}
\includegraphics[width = \linewidth]{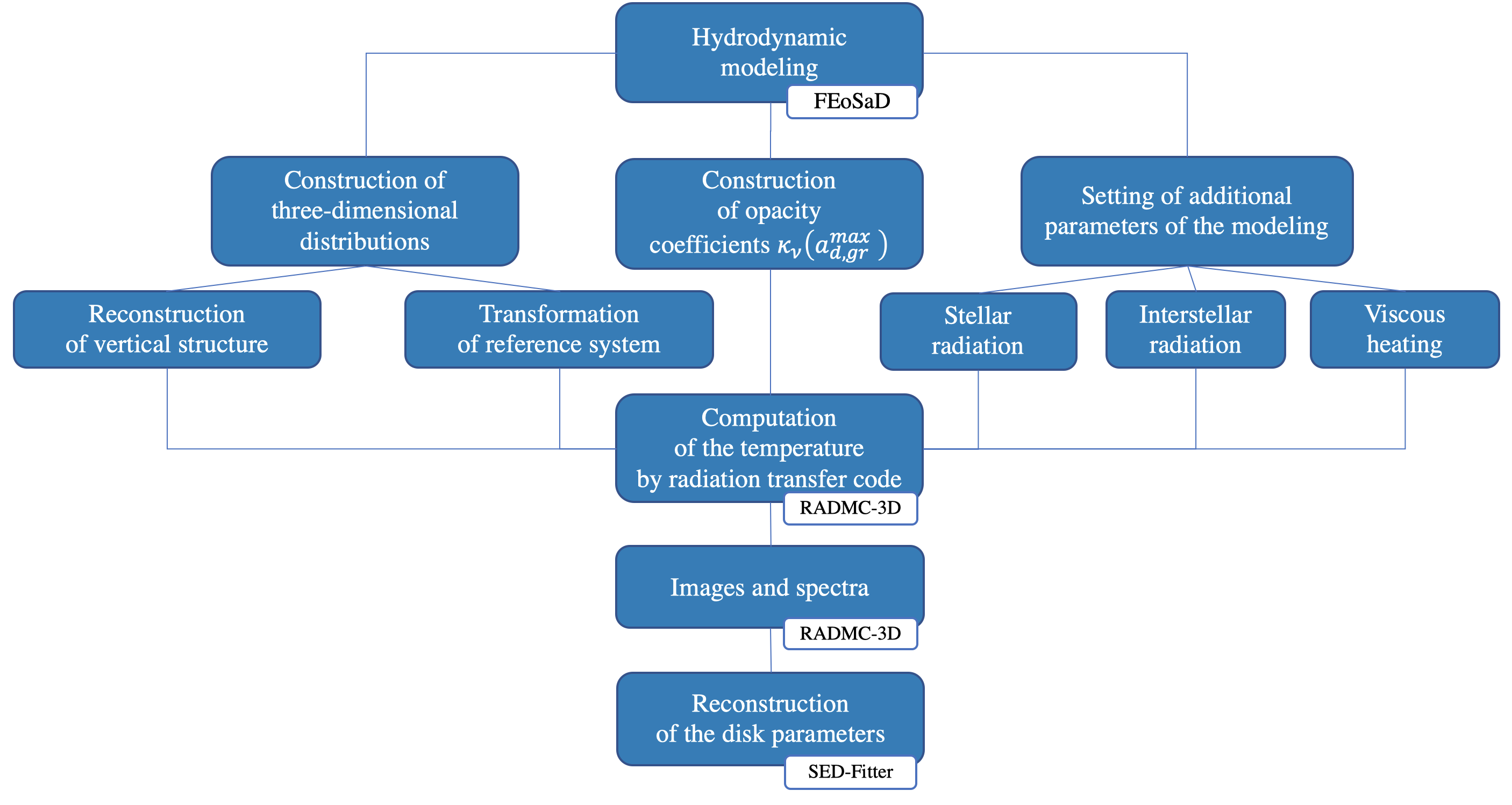}
\caption{Flow-chart describing the main stages of the modeling of synthetic images and spectra of a protoplanetary disk based on the data of hydrodynamic modeling.}
\label{fig_01}
\end{figure}

The flow-chart describing the main stages of our modeling is shown in Fig.~\ref{fig_01}. In the following sections, we will describe the steps in the sequence and discuss the obtained results.

\section{RESULTS OF THE MODELING OF THE DYNAMIC EVOLUTION OF THE DISK}

Dynamic evolution of the disk was modeled using FEOSAD software package, with a detailed description presented in~\cite{2018A&A...614A..98V}. This code is an extension of the two-dimensional ($R$,$\varphi$) model of formation and long-term evolution of a circumstellar gas-dust disk~\cite{2006ApJ...650..956V}. In particular, within the framework of this model, it is possible to reproduce the regime of episodic accretion and to explain the outbursts of fuors by the fall of gravitationally bound fragments that form in the accretion disk and migrate toward the star. The FEOSAD model is one of the most detailed among currently existent models that describe self-consistent dynamics of the dust and gas, considering a large number of key physical processes. The dust in the model is present in the form of two components --- small and grown ones. Small dust (the size below $10^{-4}$~cm) is dynamically coupled to the gas, while grown dust (larger than $10^{-4}$~cm) can drift relative to the gas. In the computation of the drift of grown dust, exchange of momentum with the gas is considered and a correct calculation of dust migration is ensured in a wide range of the Stokes numbers. The fraction of grown dust in the model can increase due to the collisions with small dust and the maximum size of the grown dust grains may change. The growth of the dust particles is limited in the model by the so-called fragmentation limit, at which collisions of the dust grains lead not to their coagulation, but to destruction. Using the dynamic FEOSAD model, the details of the dust evolution in a protoplanetary disk were obtained~\cite{2019A&A...631A...1V},  as well as the influence of the inner region of the disk on its evolution~\cite{2019A&A...627A.154V}, and the dynamics of pebbles were investigated~\cite{2020A&A...637A...5E}. 

\begin{figure}
\includegraphics[width = 0.75 \linewidth]{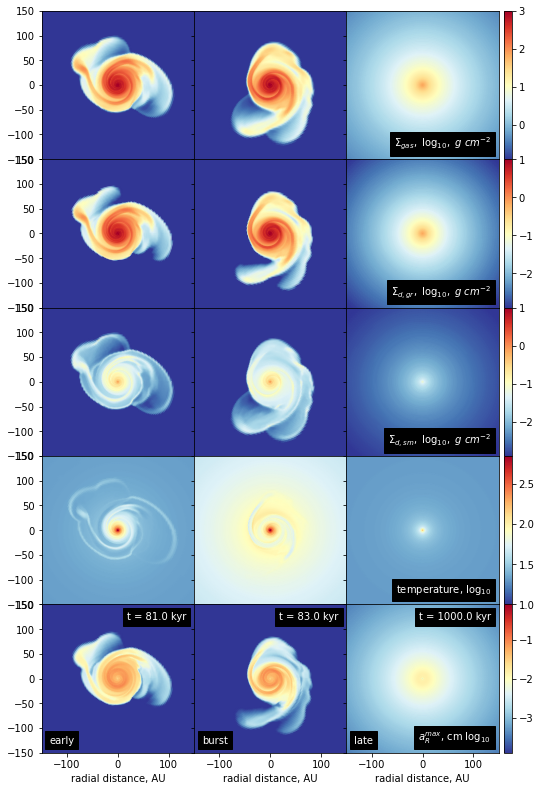}
\caption{Modeling results of the dynamic evolution of a gas-dust disk using FEOSAD hydrodynamic code. Distributions of the main parameters of the disk in the 300 $\times$ 300 AU region at different times of evolution: 81000~yr, 83000~yr, and 1~Myr are shown in the left, middle, and right columns of the panels, respectively. The upper row corresponds to the distribution of the gas surface density, the 2nd one --- to the surface density of grown dust, the 3rd row --- to the surface density of small dust, the 4th row --- to the temperatures in the equatorial plane, the lower row --- to the distribution of the maximum size of dust grains. Color scale is logarithmic}
\label{fig_02}
\end{figure}

In our calculations, the main parameters correspond to the basic model presented in~\cite{2018A&A...614A..98V}, with the exceptions listed below. The mass of the initial molecular cloud was chosen to be 0.5~$M_{\odot}$, the value of fragmentation velocity of dust was reduced to 3 m/s.
As an inner boundary condition, we used the <<smart>> cell model from~\cite{2019A&A...627A.154V} with an accretion efficiency $\xi$=0.5. Also, we applied the updated absorption and scattering coefficients for the dust population (see Section~3.2 for the details). With the parameters used, the disk forms at 30000~yr after beginning of the cloud collapse.
Three time instants were chosen for the study. In Fig.~\ref{fig_02}
we show the distributions of the main physical parameters of the disk (surface density of the gas, grown and small dust, equatorial temperature and maximum size of the grown dust) at each of the selected time instants. The left column corresponds to the age of 81000~yr, in which a young, fragmented gravitationally unstable disk with two pronounced spirals appears. The middle column of Fig.~\ref{fig_02} shows the system at the age of 83000~yr. At this instant, an outburst of accretion occurs, due to the development of the MRI. In the applied hydrodynamic model, MRI flare is activated inside the central absorbing cell when the threshold temperature of 1500 K is reached there, i.e., the temperature becomes sufficient for thermal ionization of the medium. Finally, an evolved, less dense, axisymmetric disk, which expanded during evolution, is shown in the right column. The age of this disk is 1~Myr. Regular structure and the absence of extreme conditions in such a disk are comparable to the simplest models used in the interpretation of observations.

The distributions of grown and small dust are shown in the second and third rows of Fig.~\ref{fig_02} respectively. For all selected time instants, the structure of dust disks basically repeats the structure of gaseous ones with small deviations, and the density of the grown dust is higher than that of the small dust. It should be noted that in the early stages of evolution, all dust outside the disk is small.

The fourth row of Fig.~\ref{fig_02} shows the distributions of equatorial temperature. At the age of 81000~yr, the spirals are clearly visible. Filament-like formations with an enhanced temperature are also distinguished at the distance of 50--100 AU, which corresponds to the place where the matter falls from the envelope onto the disk. Here, the elevated temperature is a consequence of adiabatic heating by compression. In a disk with an accretion burst, the temperature throughout the entire disk is higher, but the spirals can be still traced. The effect of the flare completely overcomes the effects of adiabatic heating at the interface between the envelope and the disk, so that filament-like formations are not visible. The evolved disk is cooler due to the lower accretion luminosity of the star and a quieter disk evolution.

\begin{figure}
\includegraphics[width = \linewidth]{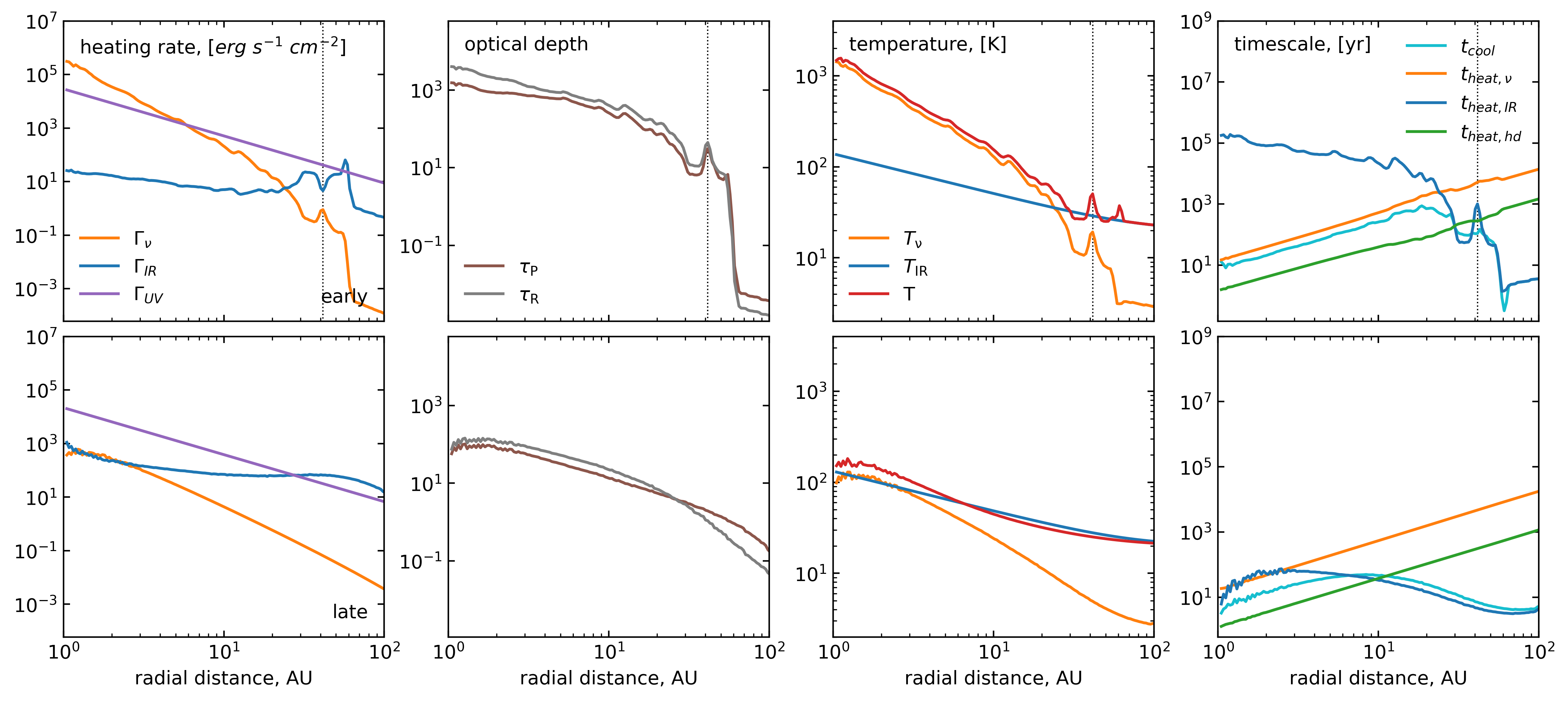}
\caption{Radial thermal structure of the gas-dust disk based on the results of the dynamic model calculation at the time of 81000~yr (the upper group of panels) and 1 Myr (the lower row). Radial cut corresponds to the angle $\varphi = \dfrac{5 \pi}{4}$, measured from the positive direction of the X-axis in Fig.~\ref{fig_02} counterclockwise. Left column: radial distributions of viscous heating (orange line), heating in the equatorial plane by IR radiation (blue line), and surface heating by UV radiation of a star (purple line). Gray and brown lines in the second column show the dependences of the vertical optical thicknesses, calculated according to Planck and Rosseland. Third column: actual temperature in the equatorial plane of the disk (red line), equilibrium temperature calculated from the condition of the equality of viscous heating and cooling (orange line), equilibrium temperature calculated from the condition of the equality of IR heating and cooling (blue line). The right column of panels shows characteristic timescales of various heating processes: IR radiation (blue line), viscosity (orange line), adiabatic compression (green line). Characteristic cooling time is shown in blue. Vertical dotted lines indicate position of the spiral in this azimuthal cut.}
\label{heating}
\end{figure}

Note, within the framework of this model, the main dust growth occurs at the early stages of the evolution. Thus, at the instants of 81000 and 83000~yr, the maximum size of the dust grains approaches $\sim$1 mm (the bottom row of Fig.~\ref{fig_02}).  Later, the maximum size of dust grains is determined by the processes of fragmentation and migration, leading to the decrease of grains size with time. For the model at the age of 1~Myr, the maximum size drops to 0.3 mm. An interesting feature is the reduced size of dust particles in the central regions of the disk. This is due to the increased temperature there, and, hence, to the reduced fragmentation barrier.
Let us consider in more detail the thermal structure of the disk along one of the radial directions in the disk for the instants of 81000~yr and 1~Myr. The left panel of Fig.~\ref{heating} shows the rate of heating of the disk surface by the UV radiation of the star $\Gamma_\text{UV}$, the rate of heating of the equatorial plane by the IR radiation of the disk itself $\Gamma_\text{IR}$, as well as the heating rate by viscous dissipation $\Gamma_{\nu}$. These rates are calculated as follows:
\begin{eqnarray}
\label{UVheat}
&& \Gamma_\text{UV} = \frac {L_*}{4 \pi r^2} \cos{\gamma_{\rm irr}} \\
\label{vischeat}
&& \Gamma_{\nu} = \frac{9}{4} \frac{GM_*}{R^3} \Sigma \nu_\text{vis} \\
\label{IRheat}
&& \Gamma_\text{IR} = \frac{8 \tau_{\rm p} \sigma T^4_{\rm irr}}{1 + 2 \tau_{\rm p} + \frac{3}{2} \tau_{\rm R} \tau_{\rm P}},
\end{eqnarray}
where $M_*, \ L_*$ are the mass and luminosity of the star, $R$ is the distance to the star, $\Sigma$ --- surface density, $\nu_{\rm vis}$ --- kinematic viscosity, 
$\tau_{\rm R}, \tau_{\rm P}$ --- Rosseland and Planck
mean optical thickness, respectively, $T_{\rm irr}$ --- radiation temperature at the surface of the disk, which is defined by the background radiation and stellar radiation incident on the disk at the angle $\gamma_{\rm irr}$ (see.~\cite{2008NewAR..52...21L} and formulae (6) and, (8) in~\cite{2017A&A...606A...5V}).

For the model with an age of 81000~yr, viscous heating dominates over the heating by radiation in the equatorial plane of the disk within the inner 20~AU, despite the fact that the stellar radiation flux at the disk surface exceeds the rate of viscous dissipation only in the region $r<5$~AU. Such a situation is due to the high values of the optical thickness of the disk respective to the thermal radiation within 20~AU: optical thickness amounts to hundreds of units (see the second panel from the left in Fig.~\ref{heating}). The energy released as a result of viscous dissipation is retained in the disk, increasing its temperature, while a high optical thickness reduces the heating rate of the equatorial disk layers by the surface layers. As a consequence, the equatorial temperature of the disk within 20~AU is determined mainly by the viscous heating, while outside 20~AU --- by stellar radiation (see the third panel in Fig.~\ref{heating}).

Selected radial cross-section intersects spiral density wave close to 40~AU, where a small jump of the equatorial temperature is observed. As can be seen from the rate of viscous dissipation, the distributions of the optical thickness, and the profile of the equilibrium temperature (calculated from the condition of equality of viscous heating and the rate of cooling), the increase of the equatorial temperature is at least partially associated with the viscous heating under conditions of a large optical thickness. It is worth noting that the heating associated with adiabatic compression of this region can also provide certain contribution to the increased temperature inside the spiral. To illustrate this, in the right panel of Fig.~\ref{heating}, we show dependence of the timescales of heating and cooling processes, as well as that of the characteristic dynamic time (the time of sound propagation in the vertical direction) on the distance to the star. Characteristic heating and
cooling times are calculated as $t = \dfrac{E}{2 G}$, where $E$ is the total thermal energy obtained from the data on the temperature and density of the gas, and $G$ is one of the sources of heating described by the Eqs.~\eqref{vischeat}, \eqref{IRheat} 
or cooling due to the emission of infrared radiation (see \cite{2017A&A...606A...5V}, formula (5)):
\begin{equation}
\Lambda = \frac{8 \tau_{\rm p} \sigma T^4_{\rm mp}}{1 + 2 \tau_{\rm p} + \frac{3}{2} \tau_{\rm R} \tau_{\rm P}},
\end{equation}
where $T_{\rm mp}$ --- is equatorial temperature. The dynamic timescale inside the spiral is shorter than the timescales of heating by IR radiation and viscosity and is comparable to the cooling timescale. Thus, if formation of a spiral is associated with the shock wave or it is accompanied by the latter, the corresponding heating can play an important role in the formation of the temperature distribution in the inhomogeneous protoplanetary disk. Note that there is also a small temperature peak at the outer side of the spiral at the radius of $\sim$60~AU. This peak corresponds to the area of the accretion of the envelope onto the disk. Here, the temperature increase is caused by adiabatic heating.

In the 1~Myr old model, optical thickness and temperature inside 40~AU decreased by about an order of magnitude compared to the model for 81000~yr. As it is seen in the left panel of Fig.~3, viscous heating is comparable to the heating by IR radiation from the surface layers only within 3~AU. Note also that the heating by IR radiation from the surface layers of the disk exceeds the stellar radiation flux in the region of 30 to 100~AU, which is associated with account of the additional contribution of the background radiation in the calculation of $\Gamma_\text{IR}$ in the FEOSAD model.

\section{CONSTRUCTION OF THE MODEL FOR COMPUTATION OF RADIATIVE TRANSFER}

We use the RADMC-3D three-dimensional code developed by C. Dullemond in order to simulate radiative transfer and to construct synthetic images of the circumstellar disk. This code is based on the Monte Carlo method and allows to calculate the radiation transfer by the dust, considering absorption, scattering, and thermal re-emission. This open access code is actively used for the modeling of PDs, circumstellar
envelopes, and molecular clouds.\footnote{www.ita.uni-heidelberg.de/~dullemond/software/radmc-3d}

\subsection{Construction of Three-Dimensional Distributions }

Results of hydrodynamic modeling are two-dimensional (vertically integrated) distributions of physical quantities. However, a complete three-dimensional disk structure is required to model the thermal structure using RADMC-3D. To form 3D distributions of gas and dust density, we restore the vertical structure of the disk for each $(R,\varphi)$-position in the disk using the formula:
\begin{equation}
\rho (z)= \rho_0 \exp \left( - \frac{z}{H} \right)^2,
\label{vstatic}
\end{equation}
where $\rho_0 = \dfrac{\Sigma}{\sqrt{2\pi} H}$ is equatorial density,
$H$ --- disk thickness, computed by FEOSAD from the condition of
vertical hydrostatic equilibrium, $\Sigma$ --- surface density for every component of the medium. Regarding the dust settling in the disk, we introduced different scale heights for gas and grown dust, $H_{g}$ and $H_{d}$ respectively.
 Here, they are related as~\cite{2001A&A...378..180K}:
\begin{equation}
H_{d} = H_{g} \left(
{
\frac{\alpha \sqrt{2}}{2} 
\left(1 + {\rm St} +
\frac{\left({(1 + {\rm St})^2 + 8\, {\rm St} ( \alpha \sqrt{2} + {\rm St})}\right)^{\frac{1}{2}}}
{(1 + {\rm St}) (\alpha \sqrt{2} + {\rm St})}
\right)
}\right)^{\frac{1}{2}},
\end{equation}
where ${\rm St}$ --- Stokes number. It is assumed in the model that the small dust is associated with the gas. Therefore, gas height scale  $H_{g}$ is used for the former. Initial temperature in the vertical direction is set uniform and equal to the equatorial one. It is important to note that the distribution~\eqref{vstatic} is hydrostatic at the initial vertically homogeneous temperature; however, after modeling thermal structure using RADMC-3D, temperature distribution throughout the disk changes and distribution ~\eqref{vstatic}
becomes non-hydrostatic. Despite this, we do not modify density distribution to facilitate more direct comparison of the results with the original data. The maximum size of the grown dust is taken the same along the vertical direction of the dust disk.

As it is shown in the previous section, the heating by viscosity can provide a significant contribution to the temperature balance of the disk. Considering RADMC-3D, we used the "distributed" source function and set the following function of internal heating (per unit volume), consistent with formula~\eqref{vischeat}, \citep{2008NewAR..52...21L}:
\begin{equation}
\Gamma_\text{vis} = \frac{9}{4} \frac{GM_*}{R^3} \rho \nu_\text{vis}
\end{equation}
Kinematic viscosity was assumed to be constant in the vertical direction and is taken from the results of the FEOSAD hydrodynamic model, where it is defined as follows:
\begin{equation}
\nu_\text{vis} = \alpha c_s H_g,
\end{equation}
where $c_s$ --- sound velocity, $\alpha=10^{-3}$. The final stage in the construction of the input distributions is interpolation of the data for a discrete grid in the spherical coordinate system used in RADMC-3D. The inner radius was chosen as 5~AU; this exceeds the inner disk size in the hydrodynamic model (1 AU). This made it possible to significantly speed up the calculations of radiation transfer. As the outer boundary of the computational domain, the radius of 250 AU was chosen.

\subsection{Obtaining Absorption and Scattering Coefficients} 
\label{section_opac}

\begin{figure}
\includegraphics[width = \linewidth]{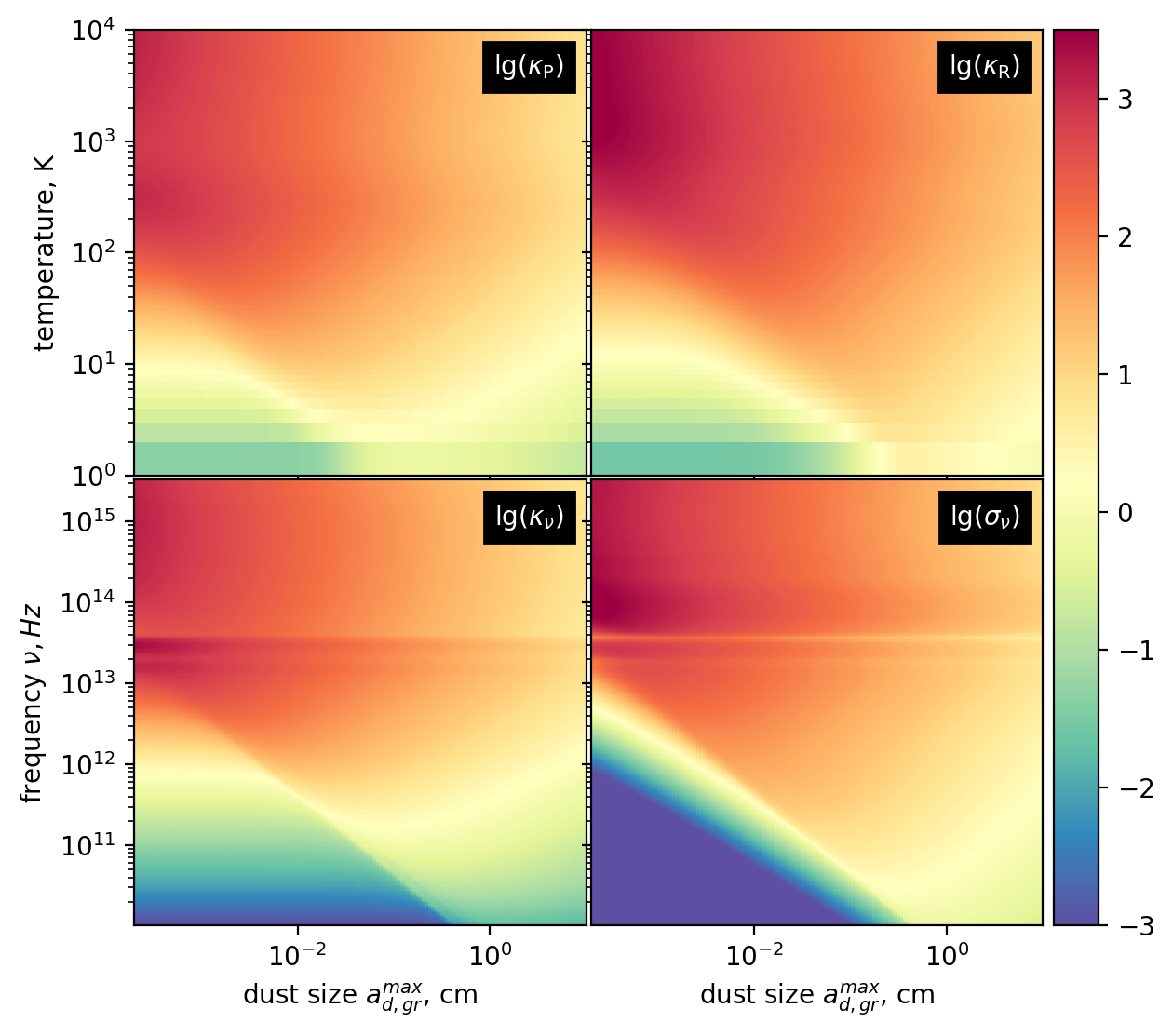}
\caption{Upper row: dependences of the Planck (left) and Rosseland (right) averaged opacities on the temperature and grains size (used in the FEOSAD dynamic model). Bottom row: dependences of the absorption (left) and scattering (right) coefficients on the frequency and size of dust grains (used to calculate the radiation transfer by RADMC-3D).}
\label{fig_04}
\end{figure}

In the original FEOSAD model~\cite{2018A&A...614A..98V} the frequencyaveraged opacities from~\cite{2003A&A...410..611S} were used. However, these opacities were calculated for a fixed size distribution of the dust. But the dust in the FEOSAD model evolves---every computational cell can have its own maximum size of dust grains. To match the dynamic and thermal dust models, we have switched to the new opacity coefficients. Using the Mie theory, we first calculated spectral absorption and scattering coefficients for different values of the maximum size of dust grains. Here, the distribution of dust grains
over size was taken as a power law $n(a)\propto a^{-3.5}$ with the minimum radius of dust grains $a_{\rm min}=5\times 10^{-7}$~cm in accordance to the dynamic model of the dust. Dust particles were assumed to contain only silicates. Frequency-dependent absorption and scattering coefficients as a function of the maximum size of dust grains are shown in Fig.~\ref{fig_04}. Note that obtained spectral absorption coefficients substantially depend on the maximum dust size, emphasizing the need for their introduction.

Next, using computed spectral absorption and scattering coefficients, Planck and Rosseland averaged opacities were computed. The dependence of the Planck $\kappa_P(T,a_{max})$ and Rosseland opacity on the temperature and the maximum size of dust grains is also shown in Fig.~\ref{fig_04}. 
As expected, two-dimensional morphology of the $\kappa_{P}(T,a_{max})$ distribution follows the morphology of the $\kappa_{\nu}(a_{max})$. distribution. When forming the averaged opacity coefficients at difference to \cite{2003A&A...410..611S}, in our computations of averaged opacities, we did not consider the possibility of dust grain evaporation at high temperatures and did not use gas opacities.

During dynamic evolution modeling of the disk frequency-averaged opacities that depend on the temperature are used, the original spectral absorption and scattering coefficients must be adopted to simulate radiative transfer by RADMC-3D. In every cell of the computational grid for RADMC-3D, depending on the maximum size of the dust grains, distributions of  $\kappa_{\nu}$ and $\sigma_{\nu}$ are specified. Thus, we use consistent dust models to simulate both disk evolution and radiative transfer.

\subsection{Set-up of Additional Modeling Parameters}

To calculate the thermal structure and observational manifestations of the disk, it is necessary to set the parameters of stellar and interstellar radiation. While the stellar radiation is the main source of disk heating (with the exception of the inner parts of the disk and fragments, where viscous heating and the work of pressure forces are also important), interstellar radiation can contribute to heating the outer regions of the disk. The parameters of the star (photosphere luminosity, mass, radius, and accretion rate) are calculated in the course of the hydrodynamic simulation of the protostellar disk itself. The data on stellar mass and radius are transmitted directly to RADMC-3D and used there. Furthermore, we assume that the star radiates as a black body. In this case, it is sufficient for RADMC-3D to provide the effective temperature of the central source as:
\begin{equation}
T_\text{eff} = \left(\frac{L_\text{phot} + L_\text{acc}}{4 \pi \sigma R^2_*}\right)^{1/4},
\end{equation}
where $L_\text{phot}$ è $L_\text{acc}$ --- photosphere and accretion luminosities, respectively, $\sigma$ --- Stephan--Boltzmann constant,  
$R_*$ --- the radius of the central star.

In addition, the disk is irradiated by the isotropic interstellar radiation. The average intensity of the latter was set as:
\begin{equation}
J_{\nu} = D \cdot B_{\nu}(T_{\rm bg}),
\end{equation}
where $D=10^{-16}$ is the dilution factor, $B_{\nu}$ is the Planck function,
$T_{\rm bg}=2\times 10^4$~K is the temperature of the interstellar radiation.

\section{RESULTS OF THE RADIATIVE TRANSFER MODELING}

Our simulations have shown that the introduction of an internal heating source into calculations of the disk temperature by RADMC-3D significantly increases computational expenses (by a factor from several dozens to hundreds compared to case of zero internal source). This increase was especially strong for disk models at the early phase of disk evolution, amounting to several months of continuous computations, which is practically unacceptable. In particular, with a zero internal source, calculation by a 72-core workstation takes 24--36 hours. If the internal source is switched on, computational time for the same number of photons increases to 150--200 days. Obviously, this is due to the high optical thicknesses of the disk in these models. As a result, we managed to simulate the thermal structure of a disk with a full set of photons (N=$10^9$) without considering viscous heating for all models, as well as the model for the age 1 Myr that considers viscous heating (due to the lower density of the evolved disk and, as a consequence, lower optical thicknesses). For the models with viscous heating at an early stage of evolution, calculations were performed
with a reduced set of photons, N=$10^7$. This made it possible to draw conclusions about the effect of viscous heating upon temperature distribution. However, for a small number of photons, temperature fluctuations are comparable to the temperature itself. Therefore, to construct images and spectral energy distributions, we will use the models without viscous heating. Next, we will sequentially analyze results of simulation of temperature and synthetic images.
\subsection{Thermal Structure}

\begin{figure}
\includegraphics[width = \linewidth]{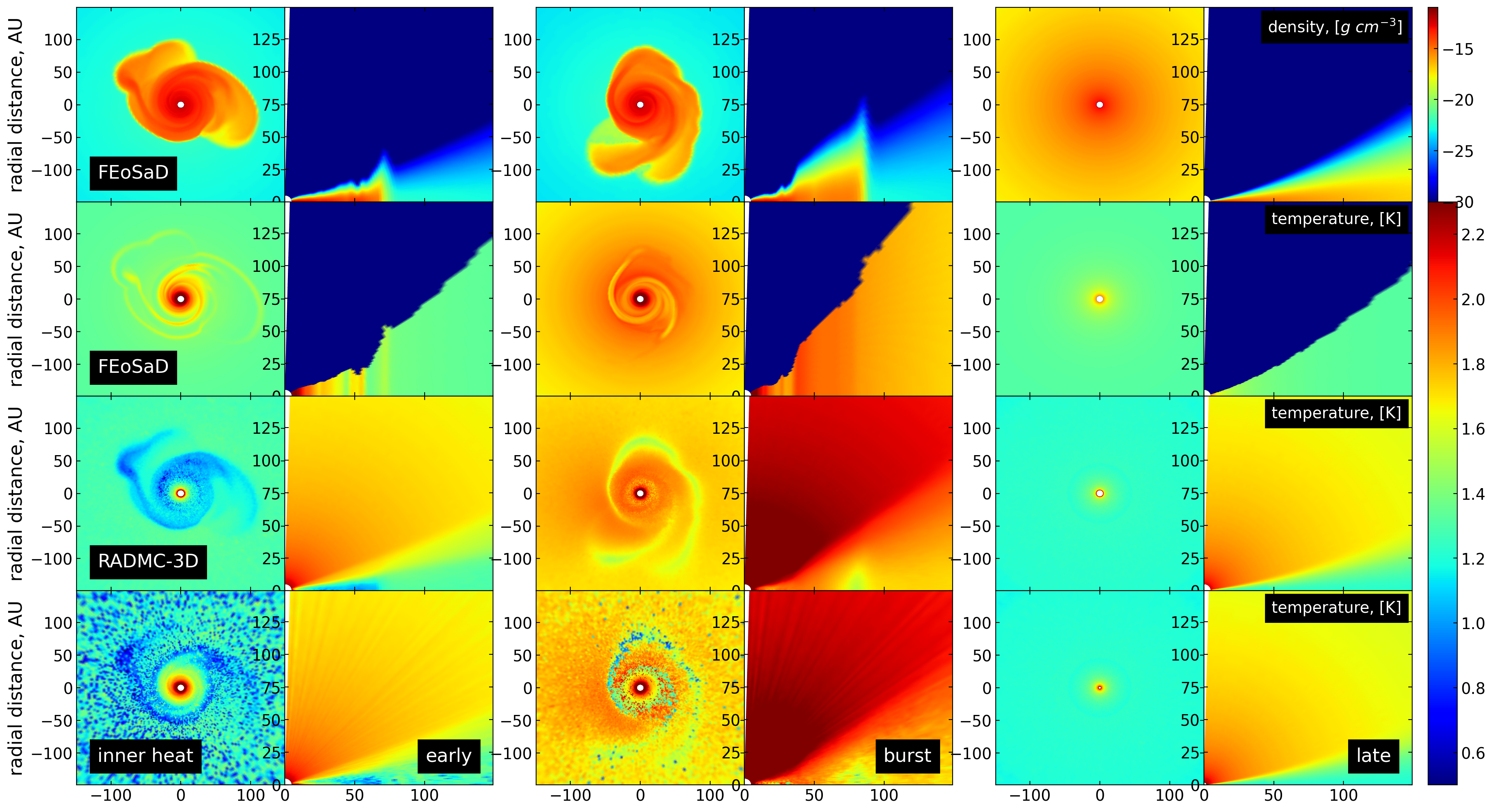}
\caption{Results of the modeling of the thermal structure of a young quiet disk at 81000 yr (left panel), a disk with a flare at 83000~yr (central panel), and an evolved disk at 1~Myr (right panel). Every panel presents density distributions (top row), FEOSAD temperature (the second row), RADMC-3D temperature with zero internal source (the third row), RADMC-3D temperature with internal heating turned on (bottom row). The left column of the panels shows the equatorial distributions; the right column of the
panels shows the meridional cut along the angle
$\varphi = \dfrac{\pi}{4}$, measured from the positive direction of the $X$-axis.}
\label{alltemp}
\end{figure}

\begin{figure}
\includegraphics[width = \linewidth]{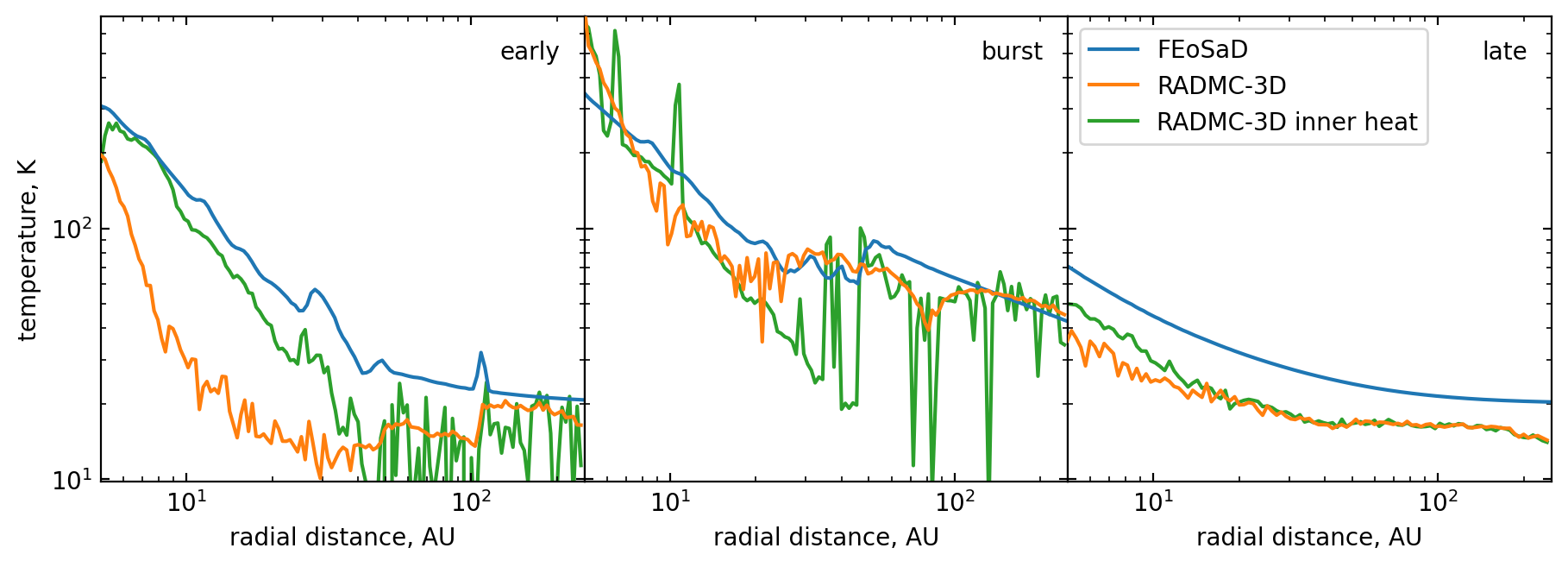}
\caption{Distributions of the equatorial temperature in the initial dynamic model (blue profiles) and after calculations by the RADMC-3D complex: orange profiles correspond to the model without viscous heating, green ones---to the model with viscous heating. The left, central, and right panels correspond to the disk ages of 81000~yr, 83000~yr, and 1~Myr.}
\label{tempcut}
\end{figure}

Figure.~\ref{alltemp} and \ref{tempcut} show calculation results of the thermal structure for the selected disk models using RADMC-3D. For the model without viscous heating with an age of 81000~yr, equatorial temperature throughout the entire disk is systematically lower than that obtained in FEOSAD. In the RADMC-3D calculation, it attains its minimum values in the vicinity of 30 AU and then there is a slight increase toward the outer regions. The reason for the slight increase of the equatorial temperature in the disk's outer region is due to the region becoming optically thin for UV radiation. As a result, direct and scattered UV radiation directly reaches the equatorial plane of the disk and determines the temperature in the latter. If viscous heating is considered in RADMC-3D, the temperature inside 30~AU rises significantly, approaching results obtained in the hydrodynamic model. Outside 30~AU, temperature distribution in the model with viscous heating shows a strong noise associated with an insufficient number of the photons. The most significant difference between the distributions obtained in RADMC-3D and dynamic model is that the spirals in the former have lower temperature than the outer regions of the disk. While in the dynamic model, the temperature inside the spirals is enhanced relative to the background (see Fig.~\ref{alltemp}). A possible reason for this is the fact that in simulation of radiative transfer using RADMC-3D, we did not consider the heating associated with the work of pressure forces (adiabatic heating). As it is shown by the analysis in Section 2, adiabatic heating can be an important factor. Another reason for the discrepancy between the results may be that taking into account three-dimensional structure of the disk and using the frequency-dependent opacities in RADMC-3D greatly expedites the escape of thermal radiation from the spirals, thereby leading to a lower equilibrium temperature in them.

In the model with a burst (83000~yr), the equatorial temperature distributions in the dynamic model and the RADMC-3D results do not differ so much. In this case, the decisive source of heating throughout the disk is heating by stellar radiation; therefore, the differences associated with internal sources of heating and interstellar heating are less pronounced. Equatorial temperature distributions obtained both by FEOSAD and RADMC-3D show a lower temperature inside the spirals, but in the RADMC-3D results, cold spirals are more pronounced. Inclusion of internal heating does not lead to significant changes of the overall picture. Temperature distribution is similar to that for the case of the zero internal source, except for some deviations of the peaks that are due to the noise caused by low photons number.

For the evolved disk (1~Myr), according to the RADMC-3D results, equatorial temperature is systematically lower than that in the original dynamic model, as in the case of the model at 81000~yr. Viscous heating in this model is considered using the full package of photons, which means that it is most consistent in comparison to other models. As can be seen, inclusion of an internal source slightly increases the temperature inside 20~AU, indicating the need to consider viscous heating even at this phase of the disk evolution. However, even considering viscous heating, the temperature according to the RADMC-3D results is systematically lower than the one in the dynamic model. Similar differences in the equatorial temperature distributions were also noted in~\citep{2017A&A...606A...5V} within the (2 + 1)-dimensional model with a non-steady method of radiation transfer treatment in the vertical direction. We believe that the systematically higher temperature throughout the disk in the hydrodynamic model is associated with the approximate character of the heating and cooling functions used in FEOSAD (see, formulae~(5) and (6) in~\citep{2017A&A...606A...5V}).

\subsection{Disk Images at Different Wavelengths}
  
\begin{figure}
\includegraphics[width = \linewidth]{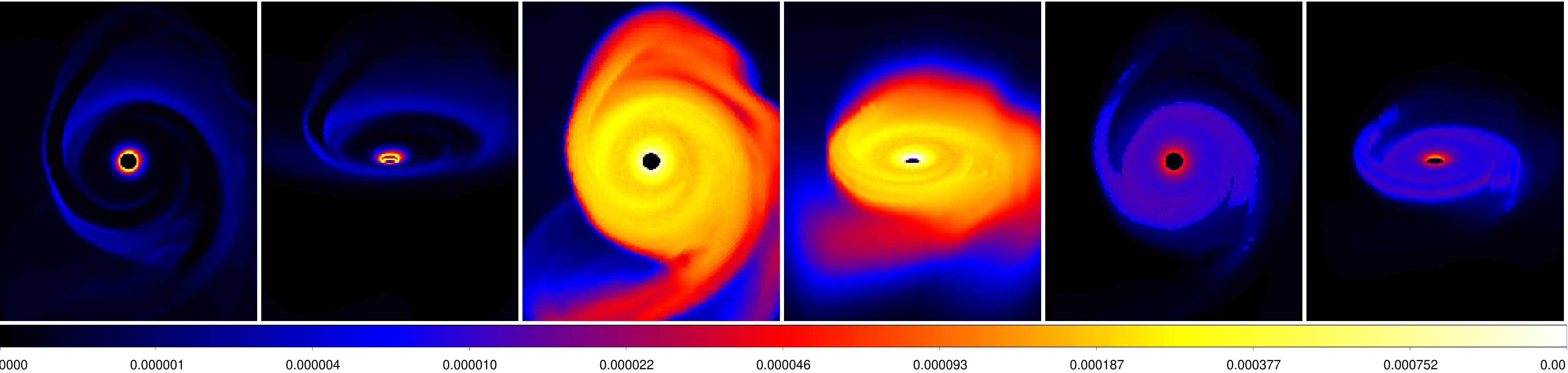}
\caption{Synthetic images of the gas-dust disk at wavelengths 1.5 $\mu m$ (left pair of panels), 300~$\mu m$ (central pair), and 1.3~mm (right pair) for the age of 81000~yr. Each pair presents a view from position with inclination 0$^\circ$ (left) and 60$^\circ$ (right) to the central axis. Intensity data are shown in color in the logarithmic scale [Jy/pixel]. The area covered by the image corresponds to the linear dimensions of $150 \times 150$ AU}.
\label{images}
\end{figure}

After calculation of the disk temperature, synthetic images were obtained at different wavelengths and with different inclinations to the line of sight, using RADMC-3D. An example of such images for a model of a young disk during an outburst ($t$ = 83000 yr) is shown in Fig.~\ref{images}. At the wavelength $\lambda$=1.3~mm (right panel in Fig.~\ref{images}) cold grown dust mainly radiates, concentrated in the equator layers of the disk, due to the settling. In this range, the disk is predominantly optically thin, so the image almost completely repeats density distribution of the grown dust. In particular, an increase of the radiation intensity near the equatorial plane of the disk can be seen in the image of the disk with an inclination of 60$^{\circ}$, indicating an enhanced concentration of the dust in this area.

At the wavelength $\lambda$=1.5$~\mu$ì (left panel of Fig.~\ref{images}) the image is defined by the scattering of the stellar radiation by small dust in the upper layers of the disk. Optical thickness of the disk in this range is large and, therefore, image morphology is determined exclusively by the surface layers of the disk. On this map, the spirals are seen as dark formations. This is due to the fact that the spirals are self-gravitating and their characteristic height scale is lower compared to the environment. As a result, spiral regions are in the shadow of the inner parts of the disk.

The image at the wavelength $\lambda$=300$~\mu m$ (central panel in Fig.~\ref{images}) is the brightest of the three ones. In this range, diffuse (thermal + scattered) radiation of grown dust from the upper layers of the disk dominates and determines a relatively uniform distribution of the brightness over the disk. It is important to note the features of the images calculated by us are close to those described in~\cite{2016ApJ...823..141D}.

\section{RECOVERING OF DISK PARAMETERS BASED ON COMPUTED SYNTHETIC IMAGES.}

For the estimate of the mass of PDs, the relation between the disk mass and emission flux in the (sub)-mm wave range is widely used~(\cite{2013ApJ...771..129A}, Eq. (2)):
\begin{equation}
\log{M_\text{disk}} = \log{F_{\nu}} + 2 \log{d} - \log{(\zeta \cdot \kappa_{\nu})} 
- \log B_{\nu}(T_\text{d}) 
\label{estimate},
\end{equation}
where $F_{\nu}$ --- is the emission flux at the frequency $\nu$, $\kappa_{\nu}$ is the absorption coefficient, $d$ is the distance to the object, $\zeta$ is the dust to gas mass ratio, and $T_\text{d}$ is the average temperature in the disk. This estimate is based
on the assumption that the disk is optically thin at a frequency $\nu$. Free parameters in this formula are the average temperature in the disk, the ratio of the mass of dust to the mass of gas, as well as the opacity coefficient. Our models show that all these parameters depend significantly on the position in the disk and the disk itself can be optically thick in its inner parts; therefore, the constraints of the formula~\eqref{estimate} are certainly not satisfied. Nevertheless, it is interesting to find out what the error of the formula~\eqref{estimate}, will be, since it is widely used for preliminary estimates. To estimate the disk mass, we used the integral radiation flux at the wavelength 1.3~mm for the following values of the parameters:
$\zeta = 10^{-2}$, $T_d$ = 20 K, $\kappa_{\nu} = 2.3\ cm^{2}\
g^{-1}$~\cite{2013ApJ...771..129A}.

A more detailed method for recovery of the disk parameters can be based on the analysis of the spectral energy distribution (SED) from the entire disk. For this, we used the SED-Fitter software package for fitting the spectra of PDs~\citep{2007ApJS..169..328R}. This complex is used to analyze observational data (see, e.g.,~\cite{2011ApJ...736..133A}). The complex is based on a phenomenological model of a disk with 14 free parameters, including stellar characteristics (mass, radius, temperature), parameters of the envelope (the rate of the infall of the matter from the envelope onto the disk, the outer radius) and the parameters of the disk itself (the mass, boundaries, accretion rate, geometric characteristics). The program searches for the most close spectra based on a pre-calculated grid of models, containing $\sim$20 thousand models and covering a wide range of stellar objects (with masses from 0.1 to 50 $M_{\odot}$) at various stages of their evolution---from the earliest ones (with the age of about several thousand years) to 10~Myr. For every model, 10 inclination angles are introduced. Thus, the total number of spectra is $\sim$200 thousand. In our calculations, we did not investigate the influence of the inclination angle on the accuracy of the parameter retrieval. Therefore, we assume that the disk axis is directed toward the observer. It is important to note in the early phases of evolution that the disk should be surrounded by an extended envelope, which obscures the disk even in the polar direction. In the modeling of the SED using RADMC-3D, we did not take this effect into account.

SED is fitted at specific wavelengths. For this, the SED-Fitter uses a system of filters. The filters can be similar to those used in the telescopes for real observations, as well as monochromatic. In this study, we use a set of monochromatic filters for wavelengths consis- tent with the set of frequencies used in the RADMC-3D calculations. However, this matching reduces somewhat the number of points involved in the fitting, since the wavelength sets used in RADMC-3D and those available in the SED-Fitter model database differ. As a result, out of 150 frequencies used in the calculation of radiative transfer and model diagrams, 70 were used for fitting. The SEDs were computed assuming the distance of 1~pc.

\begin{figure}
\includegraphics[width = \linewidth]{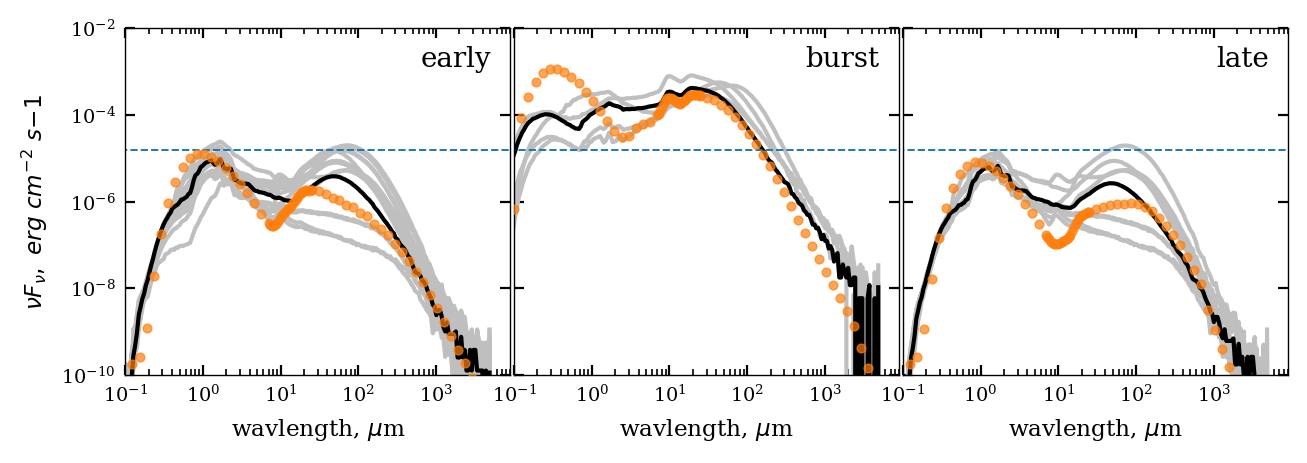}
\caption{Synthetic spectral energy distributions and the results of their fitting using "SED-Fitter" complex. Left panel shows the data corresponding to the model with the age of 81000~yr, central panel---to the 83000~yr old model, right panel corresponds to 1~Myr. Orange dots correspond to the energy distribution calculated in RADMC-3D, gray profiles show close distributions available in the model database. Solid black line is the closest distribution available in the database. The maximum flow level in the 81000~yr old model is indicated by the dashed blue line.}
\label{sedfit}
\end{figure}

In the course of fitting by the SED-Fitter algorithm, 17 to 48 close SEDs were selected for each initial spectrum (depending on the model). Results of the fitting are shown in Fig. 8. As expected, radiation flux from the disk during a flare is noticeably higher over entire spectrum. In the model of an evolved disk ($t = 1$ Myr), on average, the flux is lower than for a young, quiet disk; some flattening of the maximum toward the long-wavelengths region in comparison to the young disk is also seen. The latter effect is associated with an increase of the disk size: the disk intercepts more radiation from the star. However, the temperature of its outer regions is lower and this is reflected in the appearance of additional low-frequency radiation.

\begin{figure}
\includegraphics[width = 0.95\linewidth]{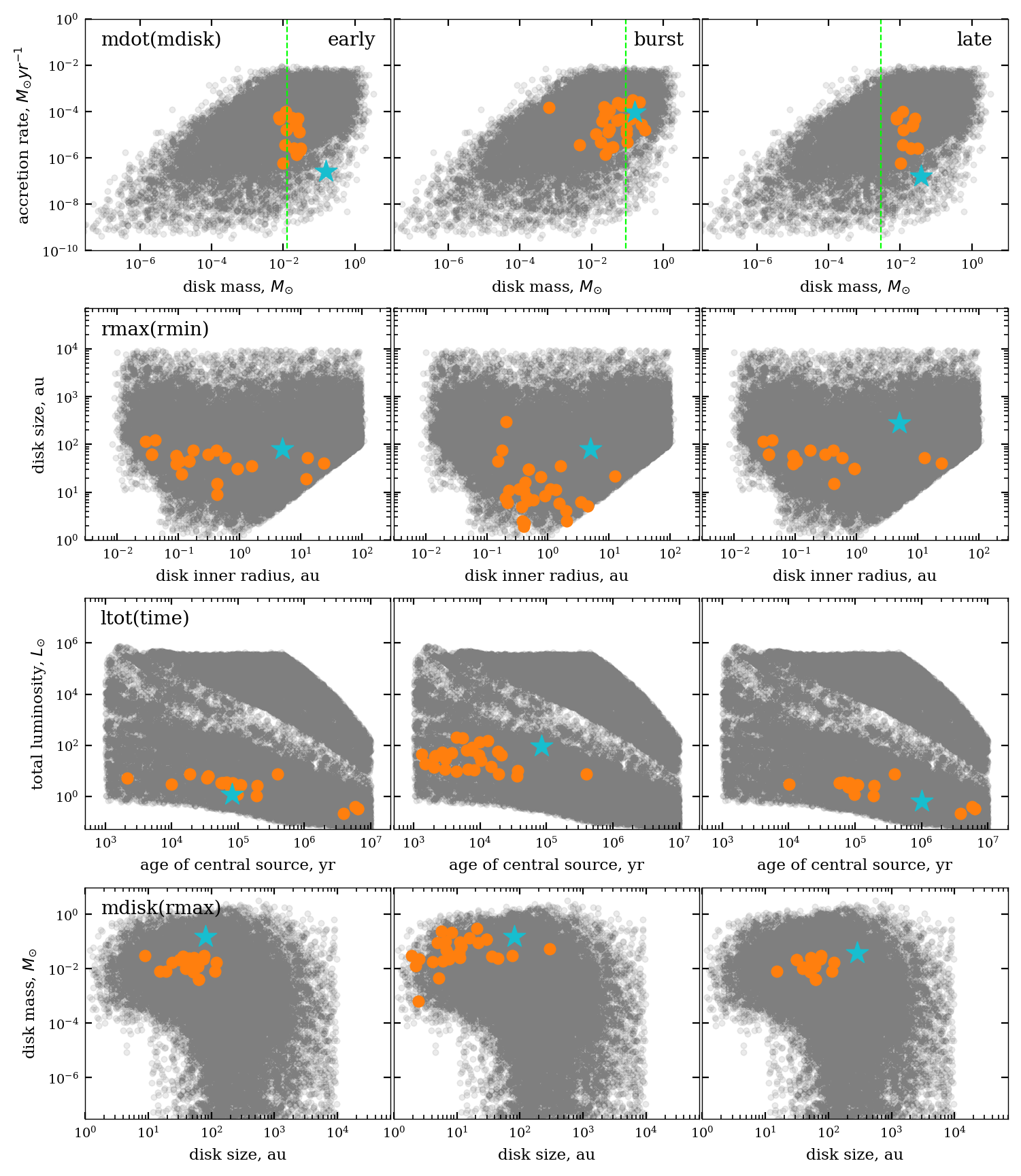}
\caption{Initial parameters of the disk model and the recovered parameters obtained by "SED-Fitter" complex. The left, middle, and right columns correspond to parameters at 81000~yr, 83000~yr, and 1~Myr. The first row shows the correspondence of parameters in the plane "accretion rate--disk mass"; the second row---in the "outer disk border--inner disk border" plane; the third row---in the "stellar luminosity--system age" plane; the fourth row---in the "mass of the disk--outer edge of the disk" plane. The models with the closest spectral energy distributions obtained during the fitting are shown in orange. Blue stars correspond to the self-consistent data obtained from the hydrodynamic simulation. Green dashed lines in the top row correspond to the masses calculated according to the expression \eqref{estimate}. Gray dots are the complete set of models in the used SED-Fitter catalog.}
\label{params}
\end{figure}

Based on results of the fitting, all 14 parameters of the system were determined, among which we identified the six most interesting to us: mass of the disk, accretion rate onto the star, the outer and inner boundaries of the disk, total luminosity of the star, and the age of the system. Recovered and original parameters of the system are shown in Fig.~\ref{params}. It can be seen that the disk masses obtained using the SED-Fitter are systematically lower than the initial values, especially in the models without an outburst ($t = 81000$~yr and $t = 1$~Myr). Also, in the models without a flare, accretion rates are several orders of magnitude higher than the initial ones. At the same time, for the model with a burst ($t = 83000$~yr), SED-Fitter reproduces well the original disk mass and accretion rate. A systematic underestimate of the disk masses was obtained not only in the fitting of the spectra, but also in the calculations of the masses using expression~\eqref{estimate}. Results of the mass calculations by formula~\eqref{estimate} are plotted in the top row of panels by green dashed lines and they are smaller than the real data. As it was already noted above, this is obviously due to the limitations of formula~\eqref{estimate}, in particular, to the assumption of an optically thin disk.

As it follows from the second row of plots in Fig.~\ref{params}, radial sizes of the disks, are recovered relatively well for the models without flash, albeit with some underestimate. At the same time, there is a significant scatter of the estimates of the disk size for the flash model, which does not allow to judge whether its determination is reliable. Restored values of the inner disk boundary in all cases have a large scatter and do not correspond to the data of the original models.

Stellar luminosity has been restored quite successfully for all disk models (see the third row in Fig.~\ref{params}). Despite small deviations, most of the best models are concentrated close enough to the real luminosity values. On the contrary, it is rather difficult to judge the age of the system based on the data obtained by the fitting of the spectra.

The bottom row of Fig.~\ref{params} illustrates the restoration results of parameters in the: "disk mass–disk size" space. These parameters can be considered as the most important observational characteristics of the protostellar disks. It can be seen that, within the SED- Fitter complex, it is not possible to restore reliably both parameters simultaneously.

Based on the results obtained, it can be argued that the SED-Fitter complex made it possible to relatively well restore only a part of the original parameters of the model protoplanetary disk. The main reasons for the discrepancies between the input and reconstructed disk parameters are obviously the differences between the disk models used to generate the spectrum and to reconstruct it. In addition, as it was noted in~\cite{2007ApJS..169..328R}, the problem of disk parameters recovery is degenerate, i.e., different features of the protoplanetary systems can manifest themselves in the SED diagrams in the same way. All this indicates the need to use spatially resolved disk observations in various spectral ranges, the analysis of which should be done on the basis of physically consistent disk models.

\section{CONCLUSIONS}
To date, the studies of PDs have made significant progress due to improved theoretical models, the development of numerical methods, and growing computing power. A huge leap has also taken place in observational methods. Without doubts, theoretical modeling and observation tools will be intensively developed in the future. However, comparison of the simulation results and observational data requires an involved work. Present study is an example of the development of a connection between a detailed dynamic model of a protoplanetary disk and its direct observational manifestations. The main results obtained in the course of this work can be distinguished as follows:
\begin{itemize}
\item The full path from the theoretical modeling of a protoplanetary system in the two-dimensional approximation of a thin disk to receiving of three-dimensional synthetic observational images and SEDs is illustrated. This will allow direct comparison of this model with observations. The features of synthesized observational manifestations of young, gravitationally unstable disks in the embedded phase, a disk during an accretion luminosity flash and an evolved disk are shown.

\item The differences in the thermal structure of the disk, obtained by hydrodynamic modeling and by calculations by the three-dimensional radiation transfer code RADMC-3D are illustrated. The contribution of various heating mechanisms in the disk is shown. In the young disks, viscous heating is an important source of heating in their optically thick inner regions. Another significant source of heating in inhomogeneous disk structures can be adiabatic heating by compression (shockwaves). However, an account of internal sources of heating in the calculation of radiative transfer using RADMC-3D leads to a manifold increase of the computational time. This leads to the need to elaborate and to use specialized algorithms for calculation of radiative transfer.

\item By simple methods used in the analysis of real observations, physical parameters of the model PDs were calculated based on their synthetic observations. For the disks under consideration, assuming that they are optically thin, they have an average temperature and a constant mass ratio of dust to gas leading to an actual underestimation of their masses. Application of the fitting of SED makes it possible to estimate relatively well individual parameters of the disk (size, luminosity, in some cases, mass). However, the problem of recovering the parameters is degenerate and this introduces a significant scatter into certain disk parameters. Therefore, it is more expedient to compare the model with observations in terms of spatially resolved observations of disks in different spectral ranges.
\end{itemize}

\section{FUNDING}
This study was supported by Russian Foundation for Basic Research (contract No. 19-32-50146). E.V. acknowledges support by the Ministry of Science and Higher Education of Russian Federation (state science assignment to Southern Federal University VnGr /2020-03-IF, 2020).

\selectlanguage{english}
\bibliography{maikbibl}

\end{document}